\documentclass[vecphys]{svmult}

\usepackage{makeidx}         
\usepackage{graphicx}        
\usepackage{multicol}        
\usepackage[bottom]{footmisc}

\makeindex             

\begin{document}

\title*{Future ASKAP Studies of the Local Volume}

\author{Lister Staveley-Smith\inst{1}}
\institute{School of Physics, University of Western Australia, Crawley,
WA 6009, Australia
\texttt{Lister.Staveley-Smith@uwa.edu.au}}
%
%
\maketitle
\abstract{The Australian SKA Pathfinder (ASKAP) will be a powerful
instrument for performing large-scale surveys of galaxies. Its
frequency range and large field of view makes it especially useful for
an all-sky survey of Local Volume galaxies, and will probably
increase the number of known galaxies closer than 10 Mpc by a factor
of two and increase, by at least an order of magnitude, the number
detected in HI. Implications for our knowledge of the HI mass
function for the very faintest galaxies and for the structure and
dynamics of the Local Volume are discussed. }

\section{Introduction}
\label{sec:intro}

The Local Volume is a key region for the study of the properties of
galaxies, including: (1) their internal structure and dynamics; (2) 
their spatial distribution and dynamics in an environment which lies in
the outskirts of a supercluster; and (3) their complete evolutionary
history, by virtue of our ability to resolve individual stars. 
The Local Volume is particularly useful for studying the faintest
galaxies and is well-served by having a wealth of accurate 
redshift-independent TRGB distances through recent surveys by the HST 
\cite{Riz07}.

As shown elsewhere in these proceedings, detailed studies of Local
Volume galaxies in the 21cm line of neutral hydrogen have been
particularly fruitful.  These have recently been rejuvenated by
surveys at other wavebands including those of {\em Spitzer} and {\em
GALEX}.  Due to the sensitivity and resolution, such surveys have
often been drawn towards the luminous galaxy population.  However, the
Local Volume also offers a unique opportunity to study the faintest
galaxies observable and many studies (e.g. {\em SINGG, LVHIS, THINGS})
have also been careful to select their samples across a range of
intrinsic luminosity.

The Square Kilometre Array (SKA) will be a radio telescope of unprecedented
power to observe galaxies in the radio continuum and in the 21cm line.
Its sensitivity will easily surpass existing telescopes and
allow Local Volume galaxies to be studied at the highest spatial resolution.
However, it's not due to come on-line for at least a decade. Nevertheless, the
next generation of so-called `SKA pathfinders' are around the corner. Their
purpose is to test SKA technologies, yet provide sufficient sensitivity
to obtain useful science and provide a valuable source of survey 
material for the SKA itself. Examples of proposed pathfinders are the Allen 
Telescope Array (ATA) and the Apertif upgrade to WSRT in the northern 
hemisphere and MeerKAT and the Australian SKA Pathfinder (ASKAP) in the 
southern hemisphere. In this brief paper, I will look at the implications
for our knowledge of the Local Volume of proposed ASKAP surveys.

\section{The Australian SKA Pathfinder (ASKAP)}

ASKAP (formerly xNTD/MIRANdA) is a so-called `1\% SKA pathfinder' although, in
reality, it will likely have a collecting area $A$, of only 0.5\% of a square 
kilometre \cite{Joh07}. However, due to its enormous field-of-view $\Omega$, 
its speed ($\propto \Omega A^2 T_{sys}^{-2}$)
will greatly exceed that of the existing 64 to 300-m class of single-dish 
radio telescopes and even the large synthesis arrays such as ATCA, GMRT, WSRT, 
and the VLA. The properties of ASKAP as listed in \cite{Joh07} are
summarized in Table~1.

\begin{table}[t]
\centering
\caption{The planned specifications of ASKAP, as listed in the expansion
option of \cite{Joh07}.}
\label{tab:1}       
%
%
\begin{tabular}{lcl}
\hline\noalign{\smallskip}
Specification & Value  & Units \\
\noalign{\smallskip}\hline\noalign{\smallskip}
Frequency range & 700--1800 & MHz \\
Number of antennas & 45 & \\
Antenna diameter & 12 & m \\
Total area &  5089 & m$^2$  \\
System Temperature $T_{sys}$ & 35 & K \\
Field of view & 30 & deg$^2$ \\
Maximum baseline & 0.4 -- 8 & km \\
Instantaneous bandwidth & 300 & MHz \\
\noalign{\smallskip}\hline
\end{tabular}
\end{table}

ASKAP is due to be located at Boolardy Station in Western Australia in
an extremely radio-quiet environment. It will therefore be able to combine
an uncluttered view of the redshifted 21-cm Universe with powerful
widefield technology, to produce fast, deep surveys which, apart from the
enormous data volumes, will be relatively straightforward to deal with in 
data reduction pipelines.

\section{A Local Volume Survey}

A natural survey to contemplate with ASKAP, and one discussed in \cite{Joh07}
is an all-sky 21cm survey. Given the specifications listed in Table~1, and 
given a year of survey time, a survey covering $2\pi$ sr will
reach an rms sensitivity of $\sim 0.26$ mJy beam$^{-1}$ in a resolution 
element of 100 kHz
(corresponding to 21 km s$^{-1}$)\footnote{The actual frequency resolution
of ASKAP will be better ($\sim~20$ kHz), but 100 kHz is appropriate for
galaxy {\em detection}.}. Such a survey will not only be able to detect 
high-mass galaxies out to slightly beyond $z=0.2$, but will also detect 
significant numbers of low-mass galaxies in the Local Volume.

\begin{figure}[t]
\centering
\includegraphics[width=8.5cm,angle=270]{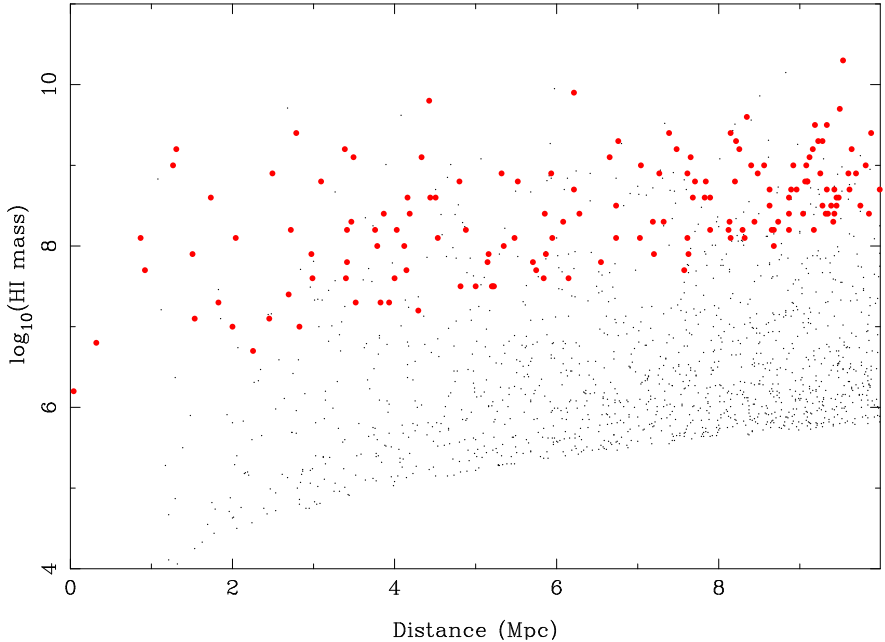}
\caption{The HI masses of 1280 putative Local Volume galaxies detected 
in a simulated 
1-yr ASKAP survey covering $2\pi$ sr are shown as small dots. Known  
galaxies, detected in the HIPASS survey \cite{Kor04} are shown as solid
circles. This simulation assumes
a compact ASKAP configuration, thus represents the maximum number of
detections expected. Local Group galaxies within 1 Mpc are not simulated.}
\label{fig:mass}
\end{figure}

Figure~\ref{fig:mass} shows a detailed simulation of the HI masses of galaxies
expected to be detected at distances within 10 Mpc by an all-sky ASKAP
survey. The simulation assumes the HIPASS HI mass function \cite{Z05},
a HIPASS mass-velocity width relationship, and a compact ASKAP configuration.
At the edge of the Local Group ($\sim 1$ Mpc), a few galaxies of HI masses 
below $10^5$ M$_{\odot}$ are expected and, at all points within the Local
Volume, galaxies down to $10^6$ M$_{\odot}$ are expected to be detected.

\begin{figure}[t]
\centering
\includegraphics[width=9cm,angle=270]{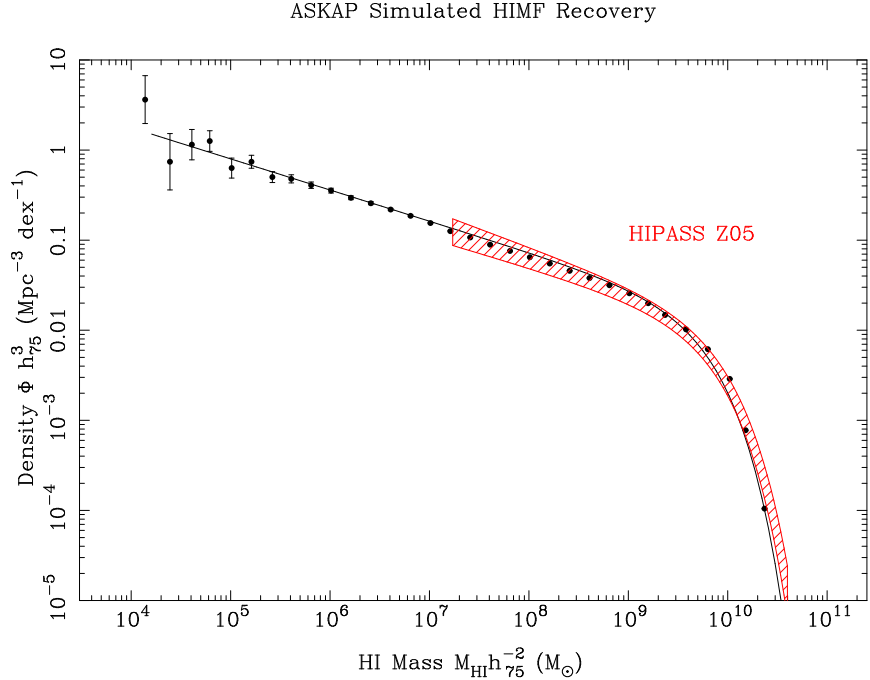}
\caption{An HI mass function derived from a simulated 1-yr ASKAP survey
covering $2\pi$ sr, again assuming a compact ASKAP configuration. For 
comparison, the known HI mass function from HIPASS \cite{Z05} is plotted in
the hatched area.}
\label{fig:HIMF}
\end{figure}

The numbers of galaxies detectable is such an ASKAP survey is enormous -- 
around $2\times 10^6$ out to the survey redshift limit. Within 10 Mpc, around
1280 are predicted, or 2560 over the whole sky if an equivalent northern
survey was feasible. This is over four times greater than the number of
galaxies ($\sim 550$) presently known to reside in the Local Volume 
\cite{K07}, and indicates the impact a future ASKAP survey is likely to make
on our knowledge of the region. However, this prediction is heavily
dependent on extrapolating the HIPASS mass function by two orders of
magnitude down the mass function! Any deviation from this has significant
implications for the prediction. This is demonstrated in Figure~\ref{fig:HIMF}
which is a mass function recovered from the above simulation. Above 
$10^7$ M$_{\odot}$ it reproduces, as it should, the HIPASS mass function.
Below that, the recovered mass function has a slope of $-1.35\pm0.01$, very 
close to the simulated slope of $-1.37$. A steeper faint-end slope (a 
natural prediction for CDM halos) will result in much greater numbers of 
low-mass objects.

\section{Discussion}

The high number of galaxies that, in all likelihood, remain to be discovered
in the Local Volume will allow a remarkably dense sampling of the
extragalactic environment of the Local Group. This will allow an accurate
mapping of the large-scale filamentary features joining the Local Group with
Sculptor and other groups. Combined with the redshift-independent distances
that are possible for such nearby objects, it will also allow a study of
the Hubble flow, infall towards filaments, and tidal stretching owing to
nearby overdense regions such as the Local Supercluster and underdense
regions such as the Local Void.

\printindex
\end{document}